\documentclass[aip,pop]{revtex4-1}
\usepackage{amsmath}
\usepackage{amssymb}
\usepackage{bm}
\usepackage{graphicx}
\usepackage{natbib}

\begin{document}


\title{A computational model for exploring particle acceleration during reconnection in macroscale systems}


\author{J. F. Drake}
\affiliation{Department of Physics, the Institute for Physical Science and Technology and the Joint Space Science Institute, University of Maryland, College Park, MD 20742}
\email{drake@umd.edu}
  \author{H. Arnold}
  \author{M. Swisdak}


\affiliation{Institute for Research in
  Electronics and Applied Physics, University of Maryland, College
  Park, MD 20742}
\author{J. T. Dahlin}

\affiliation{NASA Goddard Space Flight Center code 670, Greenbelt, MD 20771}


\begin{abstract}
A new computational model is presented suitable for exploring the
self-consistent production of energetic electrons during magnetic
reconnection in macroscale systems. The equations are based on the
recent discovery that parallel electric fields are ineffective drivers
of energetic particles during reconnection so that the kinetic scales
which control the development of such fields can be ordered out of the
equations. The resulting equations consist of a magnetohydrodynamic
(MHD) backbone with the energetic component represented by
macro-particles described by the guiding center equations. Crucially,
the energetic component feeds back on the MHD equations so that the
total energy of the MHD fluid and the energetic particles is
conserved. The equations correctly describe the firehose instability,
whose dynamics plays a key role in throttling reconnection and in
controlling the spectra of energetic particles. The results of early
tests of the model, including the propagation of Alfv\'en waves in a
system with pressure anisotropy and the growth of firehose modes,
establish that the basic algorithm is stable and produces reliable
physics results in preparation for further benchmarking with
particle-in-cell models of reconnection.

\end{abstract}

\maketitle

\section{Introduction}
\label{sec:introduction}

Observations of solar flares suggest that a large fraction of the
energy released appears as energetic electrons and ions
\citep{Lin71,Emslie04,Emslie05}. Solar observations also indicate the
highest energy electrons are closest to the inferred position of the
x-line \citep{Sui03}. In recent observations of over-the-limb flares
the limb of the sun blocked the intense emission from the
chromosphere, which enabled direct measurement of the high corona
where magnetic energy was released in the flare
\citep{Krucker10,Oka13}. The surprise was that a large fraction of the
electrons in the high emission region were in the energetic component,
indicating that most electrons in the region underwent
acceleration. Such observations are consistent with the large number
of accelerated electrons seen in flares. Further, the total pressure
of these energetic particles was comparable to that of the magnetic
field. That energetic electrons can be efficiently produced during
reconnection is not limited to flares. In {\it in situ} satellite
measurements in the distant magnetotail energetic electrons in excess
of $300$ keV were produced. They were broadly peaked around the
reconnection x-line rather than localized in boundary layers,
suggesting that electrons were able to wander over a broad region
\citep{Oieroset02}.

The observations pose significant challenges to models of electron and
ion acceleration during magnetic reconnection. These challenges
include: large numbers of electrons undergoing strong heating in
flares with the pressure of the energetic component
approaching that of the reconnecting magnetic field; the energetic
electrons peaking in a broad region around the x-line and not in
localized boundary layers; and the particle spectra exhibiting a
power law form at high energy.

These observations rule out the classical picture in which
reconnection-driven particle acceleration takes place in a boundary
layer associated with a single, large-scale reconnection site. Such a
single x-line model can not explain the large number of energetic
particles produced during reconnection nor their broad spatial
distribution. Further, reconnected magnetic field lines release most
of their energy as they expand downstream of the x-line rather than in
the diffusion regions where the topological change in magnetic
structure takes place.

On the other hand, it is also now established that current layers
typically spawn multiple magnetic islands in 2D systems \citep{Drake06,Fermo12}
or become turbulent due to the generation of multiple x-lines with
variable tilt angles in 3D systems
\citep{Schreier10,Daughton11,Liu13b,Dahlin15,Dahlin17}, especially in
the presence of strong guide fields. Observations of flux transfer events (FTEs) at the
magnetopause \citep{Russell78}, flux ropes in the magnetotail
\citep{Slavin03,Chen08} and downflowing blobs during reconnection in
the corona \citep{McKenzie99,Savage11} support the
multi-island, multi-x-line picture of reconnection. That reconnection
becomes turbulent is also consistent with recent solar flare
observations in which the production of energetic electrons was
correlated with the onset of turbulent flows \citep{Kontar17}.

Thus, observations suggest that reconnection-driven particle
acceleration takes place in a multi-island or turbulent reconnecting
environment rather than in a single, large-scale reconnection site.
To understand particle acceleration in such an environment, we write
the basic equation for the rate of energy gain of particles in a
guiding center system after summing over all particles in a local
region
\begin{equation}
\label{eqn:energy}
\frac{d W}{d t}
= E_\parallel J_\parallel 
+ \frac{P_\perp}{B} \left( \frac{\partial B }{\partial t} + {\bf v}_E \cdot {\bf \nabla} B \right) 
+ (P_\parallel+np_\parallel v_\parallel ) {\bf v}_E \cdot \boldsymbol{\kappa}
\end{equation}
where $W$ is the total kinetic energy, ${\bf v}_E=c{\bf E}\times {\bf
  B}/B^2$, $v_\parallel$ and $p_\parallel$ are the bulk parallel
velocity and momentum, and the curvature is $\boldsymbol{\kappa} =
{\bf b} \cdot {\bf \nabla} {\bf b}$ with ${\bf b}$ the unit vector
along ${\bf B}$. The parallel and perpendicular pressures are
$P_\parallel$ and $P_\perp$ and $n$ is the density. The equations
apply to any species for which the guiding-center approximation is
valid. However, for ions an additional term, the dot product of
the polarization drift into the electric field, is required since the
kinetic energy associated with the ${\bf E}\times{\bf B}$ drift is not
negligible. The first term in Eq.~(\ref{eqn:energy}) is the
acceleration by the parallel electric field. The second term
corresponds to perpendicular heating or cooling due to the
conservation of the magnetic moment $\mu$ (Betatron acceleration). The
third term drives parallel acceleration and arises from the
first-order Fermi mechanism \citep{Kliem94,Drake06a,Drake10}. Freshly
reconnected field lines downstream from a reconnecting x-line
accelerate as a result of the tension force that causes them to
``straighten''. Particles that reflect from this moving field line
receive a Fermi ``kick'' and thereby gain energy.

Betatron acceleration is typically not important during reconnection
since the release of magnetic energy leads to a reduction of $B$ and
therefore the perpendicular temperature \citep{Dahlin14}.  Depending
on the strength of the ambient guide magnetic field either
$E_\parallel$ or Fermi reflection dominates electron heating during
reconnection. Fermi reflection dominates for weak to modest guide
fields while $E_\parallel$ dominates for large guide fields. A recent
important discovery is that energetic electron production plunges in
the strong guide field limit where $E_\parallel$ dominates and
therefore $E_\parallel$ is an inefficient driver of energetic
particles \citep{Dahlin16,Dahlin17}. This result also suggests that
high frequency waves, such as double layers and electron solitary
waves, which have been identified in both observations and simulations
of reconnection \citep{Cattell05}, are not a major driver of energetic
electrons during reconnection. Importantly, in a regime where Fermi
reflection dominates, particle energy gain and magnetic energy release
are directly linked (consistent with flare observations)
\citep{Drake06,Drake13}, energetic particles spread over broad regions
and are not localized in boundary layers \citep{Dahlin15,Dahlin17},
and large numbers of particles undergo acceleration.

Energetic particle spectra in heliospheric observations typically take
the form of powerlaws. On the other hand, the particle spectra in
particle-in-cell (PIC) simulations of reconnection in the
non-relativistic regime (Alfv'en speed much smaller than the velocity
of light) typically do not form powerlaws \citep{Dahlin14,Dahlin17}
except in the limit in which the upstream plasma pressure is much
lower than that of the magnetic field (extremely low plasma $\beta$)
\citep{Ball18}. Simple ``particle-in-a-box'' models in which energy
drive and loss mechanisms are included exhibit powerlaw spectra
\citep{Drake13,Guo15}. The hardest spectra from such models have
distribution functions $f$ that scale as $v^{-5}$, which correspond to
the upper limit so that the integrated particle energy remains
finite. The particle fluxes at this limit scale as
$\varepsilon^{-1.5}$ with $\varepsilon$ the particle energy. Electron
fluxes that scale as $\varepsilon^{-1.5}$ have been observed in solar
flares \citep{Holman03}. Ion fluxes typically scale as
$\varepsilon^{-1.5}$ in the solar wind \citep{Fisk06} and in the outer
heliosphere \citep{Stone08,Decker10}. Thus, simulations of
reconnection-driven particle acceleration that are large enough to
include realistic loss mechanisms appear to be required to explain
observations.

The fundamental question is how to explore particle acceleration in
macro-scale reconnecting systems such as the solar corona where the
separation between kinetic scales and macro-scales approaches
$10^{10}$ (the Debye length is less than a centimeter for $n\sim
10^{10}/\text{cm}^3$ and $T_e\sim 100$eV while macro-scales approach
$10^4$km). The development of Parker-like transport equations that
describe reconnection-driven particle acceleration illuminate the
important physical processes that control spectra (pressure
anisotropy, feedback on the Fermi drive, particle loss versus energy
gain times)\citep{Drake13,Zank15,Montag17,Li18b}. They also yield
guidelines on the range of spectral indices that are possible in
reconnecting systems. As discussed previously, in non-relativistic
reconnection, the spectral index of the energetic particles can not
fall below $1.5$ \citep{Drake13}. However, such models are not able to
directly describe the reconnection dynamics of a given event such as
an impulsive flare in the sun's atmosphere even when they are paired
with the MHD description of the system -- scattering in such models is
assumed to be strong enough so that the energetic particles are tied
to the local fluid and so are unable to stream along ambient magnetic
field \citep{Li18b}. Such strong scattering, however, is inconsistent with
solar flare observations \citep{Kontar14}. 

Exploring the dynamics of test particles in the MHD
fields produces useful information about how particles gain energy
\citep{Onofri06,Birn04,Kowal11,Guidoni16}. However, the energy going
into the energetic particles can run away since there is no feedback
on the MHD fields. It is also possible to embed PIC models into
large-scale MHD descriptions at selected locations where reconnection
takes place \citep{Toth16}. However, such models presume that particle
energy gain is highly localized in space around isolated x-lines,
which is not consistent with the description of particle energy gain
during the development and interaction of macro-scale magnetic islands
or the development of turbulence in large-scale current layers.

The problem with conventional PIC codes in the context of modeling
large-scale systems is that the Debye length has to be resolved to
avoid non-physical heating of the electron macro-particles. Implicit
PIC models avoid this constraint but still need to resolve the
electron and ion inertial scales \citep{Lapenta06}. Conventional hybrid
codes (fluid electrons and macro-particle ions) can not model electron
acceleration and must still resolve the ion inertial scale and the ion
Larmor radius and therefore can not be used to explore energetic
particle spectra in macroscale systems.

The fundamental question is whether kinetic scale boundary layers play
an essential role in the development of particle energy gain during
impulsive flares in macro-scale systems such the sun's corona. The
rate of reconnection in kinetic descriptions corresponds to inflows
that are around $0.1V_A$ where $V_A$ is the Alfv\'en speed based on
the upstream reconnecting magnetic field
\citep{Shay99,Shay07,Karimabadi07}. On the other hand, MHD descriptions
of reconnection at low resistivity generate multiple magnetic islands
and yield reconnection rates that, while somewhat slower than in
kinetic models, are, nevertheless, insensitive to plasma resistivity
\citep{Bhattacharjee09,Cassak09,Huang10}. The inclusion of
current-driven resistivity can boost MHD reconnection rates to values
comparable to kinetic models. Kinetic boundary layers control the
regions where $E_\parallel$ is non-zero
\citep{Pritchett04,Drake05}. However, it is Fermi reflection and not
$E_\parallel$ that is the dominant driver of energetic
particles. Particle energy gain from Fermi reflection takes place over
macro-scale regions where magnetic fields are releasing energy and
takes place even where $E_\parallel=0$. Physically, particles moving
along bent field lines have curvature drifts along the reconnection
electric field and therefore gain energy as long as $\boldsymbol{\kappa}\cdot
{\bf v}_E$ is positive (see Eq.~(\ref{eqn:energy})). The conclusion
therefore is that including kinetic-scale boundary layers is not
required to describe the dynamics of energy gain of the most energetic
particles in macroscale systems. The MHD model is a reasonable
description of heating during magnetic reconnection -- either through
the formation of switch-off slow shocks in anti-parallel reconnection
or a combination of rotational discontinuities and slow shocks in the
case of reconnection with a guide field \citep{Lin93}.

We conclude therefore that we can explore particle acceleration during
magnetic reconnection in macroscale systems without resolving the
kinetic scale boundary layers that limit traditional kinetic
models. Here we present a novel computational model that combines the
MHD description of the plasma dynamics with a macroparticle
description but in which all kinetic scales are ordered out of the
system of equations. The macro-particles can be small in number
density but can contribute a pressure that can be comparable to the
pressure of the reconnecting magnetic field. They move within the MHD
grid and are advanced in parallel with the fluid equations using the
guiding center equations based on the MHD electric and magnetic
fields. The particles feed back on the MHD fluid through their
pressure-driven ${\bf J}_{h}\times {\bf B}$ force. The entire system
conserves the total energy, including that of the MHD fluid (ions and
the bulk electrons), the magnetic field and the kinetic energy of the
macro-particles. In the early phase of exploration of this model, we
are treating only electrons as macroparticles but the ions can also be
similarly treated.

There have been earlier efforts to couple the MHD equations to a
gyro-kinetic model for studying the stability of Alfv\'en waves
\citep{Cheng91} and the internal kink mode in tokamaks
\citep{Park92}. However, the gyrokinetic model orders out Fermi
reflection, which for exploring particle acceleration during
reconnection is essential.  The basic ordering that we adopt is
consistent with that discussed by Kulsrud in which Fermi reflection is
retained \citep{Kulsrud83}. Overall energy conservation was not
discussed in this previous work. Others have coupled the MHD equations
to a general kinetic particle description \citep{Bai15}. The
importance and challenge of producing a set of equations that
conserves energy exactly has been discussed previously \citep{Tronci14}.

In Sec.~\ref{sec:eqns} we present the basic equations and discuss how
the energetic component feeds back on the MHD fluid, leading to a set
of equations in which total energy is conserved. In
Sec.~\ref{sec:tests} we show the results of early tests of the code on
Alfv\'en wave propagation in a system with a finite pressure
anisotropy and firehose instability that demonstrate that there are no
fundamental computational problems associated with implementing such a
model.

\section{Basic Equations and Conservation Properties}
\label{sec:eqns}

We treat a system with three distinct classes of particles: ions of
density $n$ and temperature $T_i$, cold electrons with density $n_c$
and temperature $T_c$ and energetic electrons with density
$n_h=n-n_c$. The hot electrons will be treated as macro-particles that
are evolved through the MHD grid by the guiding center equations.
Momentum equations can be written down for each of the three species,
the ions
\begin{equation}
  m_in\frac{d\bf{v}_i}{dt}=ne{\bf E}+\frac{ne}{c}{\bf v}_i\times{\bf B}-{\bf\nabla}P_i,
  \label{eqn:mom_i}
\end{equation}
the cold electrons
\begin{equation}
  m_en_{ec}\frac{d\bf{v}_{ec}}{dt}=-n_{ec}e{\bf E}-\frac{n_{ec}e}{c}{\bf v}_{ec}\times{\bf B}-{\bf\nabla}P_{ec},
    \label{eqn:mom_c}
\end{equation}
and the hot electrons
\begin{equation}
  \frac{\partial (n_{eh}\bar{\bf{p}}_{eh})}{\partial t}=-n_{eh}e{\bf E}-\frac{n_{eh}e}{c}\bar{{\bf v}}_{eh}\times{\bf B}-{\bf\nabla}\cdot\mathbb{T}_{eh}.
    \label{eqn:mom_h}
\end{equation}
${\bf v}_i$, ${\bf v}_{ec}$ and $\bar{{\bf v}}_{eh}$ are the ion and
electron cold and hot velocities and $\bar{{\bf p}}_{eh}$ is the
average hot electron momentum (an average of the local momenta of
individual particles). The hot electron stress tensor
$\mathbb{T}_{eh}$ includes both the pressure and convective
derivatives and as a consequence the inertia term in
Eq.~(\ref{eqn:mom_h}) does not include the convective derivative. The hot
electron stress tensor is given by
\begin{equation}
  \mathbb{T}_{eh}=\int d{\bf p}_e\frac{{\bf p}_e{\bf p}_e}{\gamma_e}f
  \label{eqn:T}
  \end{equation}
with ${\bf p}_e$ the hot electron momentum with distribution $f$ and
$\gamma_e$ is the relativistic Lorentz factor. The form of
$\mathbb{T}_{eh}$ for guiding center particles and the reason for
writing the hot electron momentum equation in this form will be
clarified later.  These equations are formally exact if there are
mechanisms for maintaining the isotropy of $P_i$ and $P_{ec}$. The usual
challenge in deriving the MHD equations from the multi-fluid equations
is that the electric field and Lorentz force terms are formally larger
than the other terms in the equations. In Eq.~(\ref{eqn:mom_i}), for
example, taking $v_i \sim V_A$ and $d/dt \sim V_A/L$, the inertia term
is of order $d_i/L\ll 1$ and therefore small if the ion inertial
length $d_i=V_A/\Omega_i$ is much smaller than the system scale length
$L$. The usual procedure is then to sum the two fluid equations or in
the present case the three fluid equations, which eliminates the
electric field completely and reduces the Lorentz forces to the ${\bf
  J}\times {\bf B}/c$ force of the usual MHD equation. Since $J \sim
cB/4\pi L\sim neV_A(d_i/L)\ll neV_A$, the inertial and ${\bf J}\times
{\bf B}$ terms in the MHD equations are the same order.

In the present system we carry out the same procedure while discarding
the electron inertial terms, which are small as long as $L\gg d_e$
with $d_e$ the electron inertial length. We emphasize that we are
discarding only the inertia of the bulk flow associated with the hot
electrons and not the inertia associated with individual hot
electrons. The dominant motion of individual hot electrons in the
guiding center limit is parallel to the ambient magnetic field. The
perpendicular motion arises from ${\bf v}_{E}$ with the various
perpendicular gradient drifts being much smaller. The large parallel
velocities of the hot electrons largely cancel when summed to produce
a large parallel pressure but not a large streaming velocity. Because
we are discarding the electron fluid inertia, in summing the three
momentum equations we also discard the parallel electric field and the
parallel pressure gradient of the hot electrons. The hot electrons are
unable to couple to the MHD fluid along the ambient magnetic field
through their parallel pressure gradient. Their parallel motion is
instead controlled by the inertia of individual particles and
electromagnetic forces. They act only on the MHD fluid through their
forces perpendicular to ${\bf B}$. An extension of such a model to include a finite macroscale parallel electric field is
discussed at the end of the paper. Thus, summing the three momentum
equations yields
\begin{equation}
  \rho\frac{d\bf{v}}{dt}=\frac{1}{c}{\bf J}\times{\bf B}-{\bf\nabla}P-\left( {\bf \nabla }\cdot\mathbb{T}_{eh}\right)_\perp,
  \label{eqn:vmhd1}
\end{equation}
where we have suppressed the subscript so that ${\bf v}$ is the fluid
velocity with mass density $\rho$ and $P=P_i+P_{ec}$. The energetic
particles act on the MHD fluid through their stress tensor. It is
convenient, however, to express this force in terms of the hot
electron current $J_{ehT\perp}$ driven by the stress tensor. This
current is obtained from the hot electron momentum equation by first
subtracting the dominant current associated with ${\bf v}_E$ (which cancels that of the ions and cold electrons) from
${\bf J}_{eh}$. This yields
\begin{equation}
  {\bf J}_{ehT\perp}=\frac{c}{B}{\bf b}\times {\bf\nabla}\cdot\mathbb{T}_{eh}
  \label{eqn:jperph1}
\end{equation}
We now proceed to simplify the form of $\mathbb{T}_{eh}$ for guiding
center electrons. The stress tensor can be written in two distinct
components associated with the averaged hot electron convection and
the pressure. In the direction perpendicular to ${\bf B}$, the
dominant perpendicular motion of the hot electrons is given by ${\bf
  v}_E$ with other drifts being smaller in the ratio of the Larmor
radius to the macroscale $L$. For $v_E\sim V_A$ the inertia associated
with this perpendicular motion is negligible as long as
$m_e/m_i\ll\beta_{eh\perp}\sim 1$. In this limit the stress tensor
takes the usual gyrotropic form
\begin{equation}
  \mathbb{T}_{eh}=T_{eh\parallel}{\bf bb}+P_{eh\perp}(\mathbb{I}-{\bf bb}),
  \label{eqn:T_h}
\end{equation}
where $\mathbb{I}$ is the unit tensor,
$T_{eh\parallel}$ is the stress tensor along the magnetic
field ${\bf B}$ and $P_{eh\perp}$ is the usual perpendicular pressure,
\begin{equation}
  P_{eh\perp}=\int d{\bf p}_{e}\frac{p_{e\perp}^2}{\gamma_e}f,
    \label{eqn:Pperp}
\end{equation}
where in the frame drifting with ${\bf v}_E$, $f=f({\bf
  x},p_{e\parallel}, p_{e\perp},t)$ since there is no other mean drift
perpendicular to ${\bf B}$. $T_{eh\parallel}$ includes the mean parallel drifts of the hot
  electrons and can be written as a combination of the usual parallel
  pressure $P_{eh\parallel}$ plus the mean parallel convection terms,
 \begin{equation}
  T_{eh\parallel}=\int d{\bf p}_{e}\frac{p_{e\parallel}^2}{\gamma_e}f=P_{eh\parallel}+n_{eh}\bar{p}_{eh\parallel}\bar{v}_{eh\parallel}
  \label{eqn:Tpar}
\end{equation}
with 
  \begin{equation}
    P_{eh\parallel}=\int d{\bf p}_{e}(p_{e\parallel}-\bar{p}_{eh\parallel})\left( \frac{p_{e\parallel}}{\gamma_e}-\bar{v}_{eh\parallel}\right) f.
    \label{eqn:Ppar}
    \end{equation}
The hot electron parallel bulk streaming terms in
  Eq.~(\ref{eqn:Tpar}) are nominally much smaller than the parallel
  pressure since
\begin{equation}
  n_h\bar{p}_{eh\parallel}\bar{v}_{eh\parallel}\sim\frac{m_eJ_\parallel^2}{ne^2}\sim\frac{B^2}{4\pi}\frac{d_e^2}{L^2}\sim P_{eh\parallel}\frac{d_e^2}{L^2}\ll P_{eh\parallel}.
\end{equation}
 On the other hand, we demonstate below that exact energy conservation
 requires that this nominally small contribution to $T_{eh\parallel}$
 be retained since these contributions appear in the expression for electron energy gain given in Eq.~(\ref{eqn:energy}). With the form of the stress tensor given in Eq.~(\ref{eqn:T_h}), the hot electron current can be expressed as \citep{Parker65}
\begin{equation}
  {\bf J}_{ehT\perp}=\frac{c}{B}(T_{eh\parallel}-P_{eh\perp}){\bf b}\times\boldsymbol{\kappa}+\frac{c}{B}{\bf b}\times {\bf\nabla}P_{eh\perp}.
  \label{eqn:jperph2}
\end{equation}
An equivalent form for the hot electron current is
\begin{equation}
  {\bf J}_{ehT\perp}=\frac{c}{B}{\bf b}\times\left( P_{eh\perp} {\bf\nabla}\ln (B)+T_{eh\parallel}\boldsymbol{\kappa}\right) -c\left( {\bf\nabla}\times\frac{P_{eh\perp}{\bf b}}{B}\right)_\perp,
  \label{eqn:jperph3}
\end{equation}
where the first term on the right is the gradient $B$ drift, the
second is the curvature drift and the third is the magnetization
current \citep{Li18a}. The MHD equation with energetic electron feedback can then be
written as
\begin{equation}
  \rho\frac{d{\bf v}}{dt}=\frac{1}{c}{\bf J}\times{\bf B}-{\bf \nabla }P-\frac{1}{c}{\bf J}_{ehT\perp}\times {\bf B}.
  \label{eqn:vmhd}
\end{equation} 
 The calculations leading to Ohm's law in this three species system parallel that of the electron-ion system. As discussed previously, the dominant terms in Eqs.~(\ref{eqn:mom_c})-(\ref{eqn:mom_h}) are the electric field and Lorentz terms. Adding the two electron equations and discarding the pressures and stress tensor, we obtain
 \begin{equation}
   {\bf E}=\frac{1}{nc}(n_{ec}{\bf v}_{ec}+n_{eh}{\bf v}_{eh})\times {\bf B}= \frac{1}{nec}{\bf J}\times {\bf B}-\frac{1}{c}{\bf v}\times {\bf B}\simeq -\frac{1}{c}{\bf v}\times {\bf B},
   \label{eqn:ohm}
 \end{equation}
 where we have added and subtracted $n{\bf v}$ in the Lorentz force
 and again used the fact that $J\ll nev$ to eliminate the ${\bf
   J}\times {\bf B}$ or Hall term in Ohm's law. Thus, Ohm's law, which
 determines ${\bf E}$ in terms of ${\bf v}$ is unchanged from the
 usual MHD prescription. The equations for the pressure $P$ and mass
 density $\rho$ are also unchanged.

 The model is completed by the guiding-center equations for the hot electrons \citep{Northrop63}
\begin{equation}
  \frac{d}{dt}p_{e\parallel}=p_{e\parallel}{\bf v}_E\cdot\boldsymbol{\kappa}-\frac{\mu_e}{\gamma_e}{\bf b}\cdot {\bf\nabla}B
  \label{eqn:pepar}
\end{equation}
with $p_{e\parallel}$ the parallel momentum of a macroparticle
electron with its magnetic moment given by
\begin{equation}
  \mu_e=p_{e\perp}^2/2B.
  \label{eqn:mu}
  \end{equation}
 $p_{e\perp}$ is determined from the conservation of $\mu_e$. The
particle velocity is given by ${\bf v}_E$ and the parallel streaming
$v_{eh\parallel}=p_{eh\parallel}/(\gamma_em_e)$ along ${\bf B}$, the
curvature and gradient $B$ drifts being smaller in the ratio of the
Larmor radius to the macroscale $L$. The ordering of the hot electron
drifts and their energy gain in Eqs.~(\ref{eqn:pepar})-(\ref{eqn:mu})
are equivalent to Kulsrud's guiding center description
\citep{Kulsrud83}.  A critical goal in developing a credible set of
equations to describe particle acceleration is to establish energy
conservation. By taking the dot product of Eq.~(\ref{eqn:vmhd}) with
${\bf v}$ and integrating over space the energy conservation relation
takes the form
\begin{equation}
  \frac{d}{dt}W_{MHD}=-\int d{\bf x}\ {\bf J}_h\cdot {\bf E}=-\frac{d}{dt}W_h\\ =-\int
  d{\bf x}\left[ T_{eh\parallel}{\bf v}_E\cdot\boldsymbol{\kappa}+\frac{P_{eh\perp}}{B}\left( \frac{\partial B}{\partial t}+{\bf v}_E\cdot {\bf\nabla} B\right) \right]
  \label{eqn:energytot}
\end{equation}
where $W_{MHD}$ is the usual energy in the MHD description, including
the kinetic energy of the bulk flow, the thermal energy and magnetic
energy. $dW_h/dt$ is the rate of change of the energy of the hot
electrons. $dW_h/dt$ in Eq.~(\ref{eqn:energytot}) is equal to the
spatial integral of the rate of energy gain in
Eq.~(\ref{eqn:energy}). We again note that the convective terms in the
curvature in Eq.~(\ref{eqn:energytot}) are nominally small since
$d_e^2/L^2\ll 1$ but must be retained so that the energy gain in
Eq.~(\ref{eqn:energy}), which follows from Eq.~(\ref{eqn:pepar}) and the
conservation of $\mu_e$, matches that in
Eq.~(\ref{eqn:energytot}). Having equations that exactly conserve
energy facilitates testing the model and is desirable
\citep{Tronci14}. 

The equations presented above provide a complete self-consistent
system for exploring the production of energetic electrons in
macroscale systems. Since the electrons are evolved in the fields from
the MHD equations, the artificial heating associated with the PIC
model when the Debye length is not resolved is not an issue. Similar
equations can be written down that also include energetic ions
although the neglect of their inertia requires that their number
density be small. Beyond energy conservation, an important
consideration is whether the equations properly describe the feedback
of the energetic component on the MHD fluid. It is straightforward to
show that the inclusion of an ambient pressure anisotropy in the hot
component through $\mathbb{T}_{eh}$ yields the correct firehose
stability criterion. In the case of magnetic reconnection the firehose
stability boundary plays an important role in throttling reconnection
\citep{Drake06a,Drake10} and in controlling the spectral index of the
energetic particles resulting from reconnection \citep{Drake13}. The
firehose stability boundary will act similarly in this model if the
pressure in the energetic component is too high. With these equations
the production of energetic particles in realistic macroscale systems
can be explored where realistic losses can be included and the
realistic spectra of synchrotron emission from the volume and
Bremsstrahlung emission at system boundaries can be calculated for
direct comparison with X-ray observations from satellite missions such
as Ramaty High Energy Solar Spectroscopic Imager (RHESSI) and
ground-based radio observatories such as the Nobeyama Radioheliograph
(NoRH) \citep{Nakajima94} or the Extended Owens Valley Solar Array
(EOVSA) \citep{Gary18}.

\section{Tests of the {\it kglobal} model}
\label{sec:tests}
As discussed in the previous section, the pressure anisotropy of the
energetic electrons plays an important role in throttling magnetic
reconnection and limiting the energy gain of those particles
\citep{Drake06a,Drake10,Drake13}. Thus to ensure the model correctly
describes the impact of pressure anisotropy on magnetic field dynamics
we benchmark the code with two simple wave modes that are evolved in a
system with an imposed initial pressure anisotropy: the linear
propagation of stable, circularly polarized Alfv\'en waves; and the
linear growth of firehose modes. The correct solutions of both of
these tests are, of course, well known \citep{Parker58}.

The new computational model was constructed by merging the fluid
evolution equations of the {\it f3D} code \citep{Shay03} (with the
Hall terms in Ohm's law removed) and the particle treatment in the
{\it p3d} code \citep{Zeiler02}, modified to step the particles in the
guiding center limit. Time stepping is with a second order trapezoidal
leapfrog scheme with a fourth order viscosity added to each of the
fluid equations to prevent the buildup of noise at the grid scale.

In this new model the magnetic field strength, $B_0$, and density,
$n_0$, define the Alfv\'en speed, $V_{A}=\sqrt{B_0^2/4\pi
  m_in_0}$. Since there are no kinetic scales that enter the
equations, lengths and times are normalized to a macroscale length,
$L$, and Alfv\'en crossing time, $\tau_A=L/V_A$. This normalization
allows us to set the physical distance of the longest dimension in our
simulations to $2\pi L$ where $L$ can be any macroscopic scale
length. Electric fields and temperatures are normalized to
$V_{A}B_0/c$ and $m_iV_{A}^2$, respectively.  A fourth order
hyperviscosity, $\nu\nabla^4$, is included for every quantity evolved
on the grid (magnetic field, ion density, momentum and pressure and
cold electron pressure).

The tests were carried out in a system with two space dimensions with
$B_x=B_0$. The ion to hot electron mass ratio is set to $25$ (the cold
electrons are massless). For a given hot electron pressure and density
the mass ratio controls the streaming velocity of electrons through
the system. For linear waves with an imposed initial pressure
anisotropy the evolution of the pressure does not enter the equations
so the value of electron mass does not influence the dynamics. The
temperature of the ions and the cold electrons was $1/12$. For the hot
electrons, the temperature was varied to control the magnitude of the
anisotropy of their pressure tensor. The box size was varied from 256 x 64
cells to 512 x 256 cells and there were 160-320 particles per grid
cell.

\begin{figure}
\centering
\includegraphics[keepaspectratio,width=5.0in]{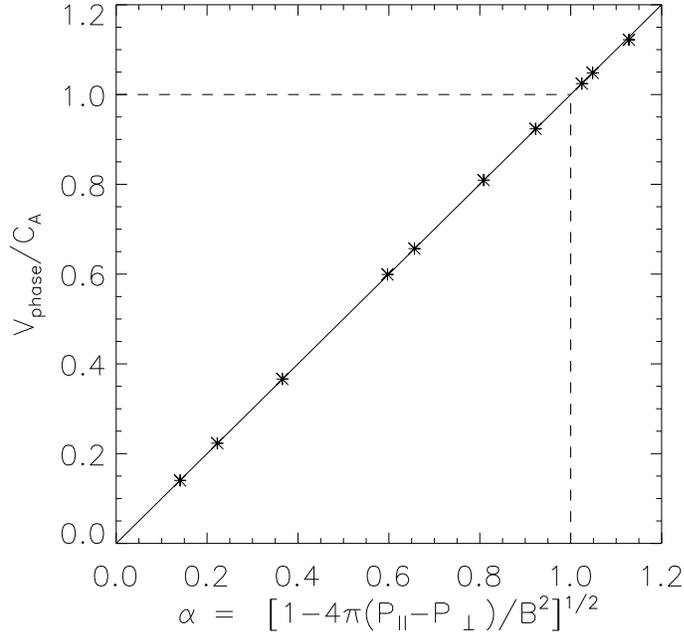}
\caption{For each of ten simulations we plot the measured phase speed
  of the Alfv\'en wave $V_p$ versus the anisotropy parameter
  $\alpha$. The solid line is what we expect from our model, which is
  the same as that of the linearized CGL equations. The dotted lines
  show where the isotropic Alfv\'en wave lies and separates the region
  where $P_\parallel$ is greater than $P_\perp$ from where it is
  smaller.}
\label{alfvenwave}
\end{figure}

In the first benchmark of the model we propagated a circularly-polarized
Alfv\'en wave along a magnetic field in a system with an imposed hot electron
pressure anisotropy. We initialized the simulations with a
perturbation with a wavelength equal to the size of the box. After
propagating the wave for a time $\tau_A$, we measured its speed. Our
equations yield the phase speed $V_p$ of an Alfv\'en wave:
\begin{equation}
V_p = V_A \sqrt{1-4\pi \frac{ P_{\parallel}- P_{\perp}}{B^2}}\equiv V_A \alpha,
\end{equation}
where $\alpha=\sqrt{1-4\pi (P_{\parallel}- P_{\perp})/B^2}$. This
result is identical to that from the Chew-Goldberger-Low (CGL) equations since in the linear
limit of the system the pressure remains unperturbed.  In Figure
\ref{alfvenwave} the wave phase speed $V_p$ is plotted as a function
of the anisotopy parameter $\alpha$. The agreement with linear wave
theory is excellent.

\begin{figure}
\centering
\includegraphics[width=27pc]{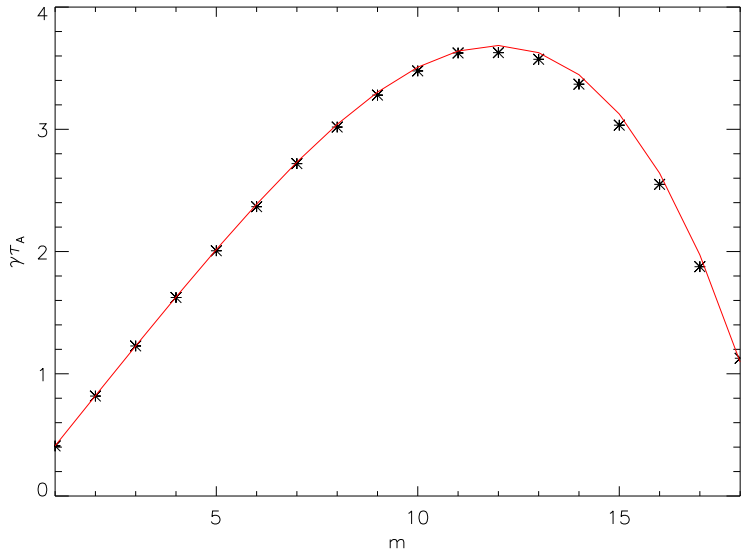}
\caption{Normalized growth rate, $\gamma\tau_A$, verus the mode number,
  $m=2\pi kL$, for a range of unstable values of $m$. The numerically
  determined values of $\gamma\tau_A$ are marked with black stars and the
  theoretical growth rate as a red line.}
\label{GrowthRate}
\end{figure}

In our second benchmark we explored the linear growth of the firehose
instability with an imposed initial unstable pressure anisotropy with
$\alpha^2=-0.16$. We initialized the simulation with small sinusoidal
perturbations for 18 values of the wavenumber, $k=m/2\pi L$, where
$m=1$, $2$, $...$ is the mode number, and the viscosity was
$\nu=6.0 \cdot 10^{-5}$. The theoretical growth rate is given by
$\gamma = kV_A |\alpha|-\nu k^4$. The viscosity controls the cutoff of
the instability at short spatial scales.  In Figure \ref{GrowthRate}
we plot the theoretical (solid red line) and numerical growth rates
(black stars) for the range of unstable wave numbers. For $m>18$ the modes are
stable. There is excellent agreement between the new model and what
one would expect from linear theory. 

\section{Summary and Discussion}
The enormous separation between kinetic scales (the Debye length, the
electron and ion inertial scales and Larmor radii) in the solar corona
(as small as a centimeter) and the energy release scales (~$10^4$km),
mean that modeling the release of energy in flares in the solar corona
and other astrophysical systems using a PIC model, which needs to
resolve the Debye scale, is not feasible even with projected increases
in computational power. Recent advances in our understanding of the
mechanisms for particle acceleration \citep{Dahlin15,Dahlin17},
suggest that these boundary layers, which control the structure of
parallel electric fields, play only a minor role in the production of
the most energetic particles. Particle acceleration is controlled by
the large-scale dynamics of magnetic fields through the merging of
magnetic islands in 2D systems and the turbulent interactions of
x-lines in the more physically realistic 3D systems. We have presented
here a new model in which we have ordered out all of the relevant
kinetic boundary layers. The result is a model that is scale
independent and therefore capable of modeling macroscale systems.

The model consists of an MHD backbone in which macroparticles
(electrons) move through the MHD grid using the guiding center
equations with electric and magnetic fields given by the usual MHD
prescription. Importantly, the energetic electrons feed back on the
MHD fluid through the perpendicular currents associated with their
anisotropic stress tensor. The consequence is that energy is conserved
exactly. Further the development of pressure anisotropy of the
energetic component (with $P_\parallel > P_\perp$) properly describes
the reduction in magnetic tension that drives reconnection and
therefore controls the feedback of the energetic particles on the
dynamics of reconnection. The equations describing the full system
consist of Eqns.~(\ref{eqn:T_h})-(\ref{eqn:mu}) with the energy
conservation relation given in Eqn.~(\ref{eqn:energytot}).  A code has
been developed to solve these equations by merging the basic
algorithms of the {\it f3D} Hall MHD and {\it p3d} PIC codes. The
resulting model has been benchmarked with the propagation of Alfv\'en
waves and firehose modes in a system with a specified initial
pressure anisotropy.

Our plans are to proceed with the exploration of electron heating and
acceleration in a simple 2D reconnecting sytem. There are a variety of
simulation results already in the literature on the scaling of
reconnection-driven, electron heating with the upstream parameters of
the system \citep{Shay14,Haggerty15}. The results of these simulations
can be compared with the results from the present model.

Before carrying out these reconnection simulations, however, we plan
to carry out an upgrade of the model to include a large-scale
parallel electric field. We have argued that the parallel electric
fields that develop in the boundary layers that form during
reconnection are not important for the production of the most
energetic particles since these boundary layers occupy very little
volume in a macroscale system -- their widths scale with the electron
skin depth. However, it has now been established that large-scale, parallel
electric fields can develop as a result of electron pressure gradients
in reconnecting systems \citep{Egedal08,Egedal12,Haggerty15}. Both electrons and
ions are heated as they enter the reconnection exhaust. Because the
thermal motion of electrons is so much greater than that of the ions,
especially for mass-ratios that approach realistic values, electrons
try to escape on the reconnected field lines threading the exhaust,
which extend into the upstream plasma that has not yet entered the
exhaust. Charge neutrality, of course, prevents the electrons from
streaming upstream and the result is a parallel potential that traps
electrons in the exhaust. This potential is not large enough to
significantly impact the most energetic electrons in the
system. However, electrons that first enter the exhaust drop down the
potential and boost their parallel velocity. This energy increase facilitates
subsequent energy gain through Fermi reflection
\citep{Haggerty15,Egedal12}. The parallel electric field
associated with the charge neutrality constraint can be calculated
from the electron parallel force balance, obtained from the sum of the electron momentum equations ((\ref{eqn:mom_c}) and
(\ref{eqn:mom_h})) projected along the magnetic field direction with
the total inertia of the electrons neglected \citep{Haggerty15}. The resulting expression for $E_\parallel$ is given by
\begin{equation}
  E_\parallel=-\frac{1}{ne}\left( {\bf b}\cdot {\bf\nabla}P_{ec}+{\bf B}\cdot {\bf\nabla}\frac{m_en_{ec}v_{ec\parallel}^2}{B}+ {\bf b}\cdot ({\bf\nabla}\cdot\mathbb{T}_{eh})\right).
  \label{eqn:epar}
\end{equation}
Note that the individual streaming velocities of the cold and hot
electrons and their associated inertias could be large but the
constraint on the total parallel current requires that the sum of the
streaming velocities be small. This is a traditional return current
picture in which hot electrons stream outwards from a region where
magnetic energy is being released but drive a return current of cold
electrons that eliminates the net electron current and prevents charge
separation of the two species. The physics argument leading to
Eq.~(\ref{eqn:epar}) is similar to that presented by Kulsrud to
calculate $E_\parallel$ \citep{Kulsrud83}. He argued that the parallel
electic field would develop to maintain charge neutrality in the
system. His expression for $E_\parallel$ includes corrections
associated with ion dynamics, which are of order $m_e/m_i$ smaller
than those retained in Eq.~(\ref{eqn:epar}). In our model the Debye
length is ordered out so the system must remain charge neutral. The
ion density is calculated with a standard continuity equation with a
velocity given by the MHD momentum equation. The energetic electron
density is calculated by mapping the energetic electrons onto the MHD
grid with an appropriate interpolation scheme. The cold electron
density is then calculated by requiring that the sum of the cold and
hot electron densities match that of the ions. The physics leading to
charge neutrality is the strong parallel motion of the cold electrons
that fills in for the hot electrons motion along the ambient magnetic
field.

Thus, our goal is to extend the present model by incorporating the
parallel electric field into the equations and then to proceed with a
comparison of electron heating in simple 2D reconnecting systems using
the new model and standard PIC.
\section{Acknowledgements}
This work has been supported by NSF Grant Nos. PHY1805829 and
PHY1500460, NASA Grant Nos.  NNX14AC78G and NNX17AG27G and the FIELDS
team of the Parker Solar Probe (NASA contract
NNN06AA01C). J.T.D. acknowledges support from the NASA LWS Jack Eddy
Fellowship administered by the University Corporation for Atmospheric
Research in Boulder, Colorado. Simulations were carried out at the National
Energy Research Scientific Computing Center. We acknowledge
informative discussions with Dr.\ William Daughton, Prof.\ A.\ B.\ Hassam and
Dr.\ G. Hammett. Simulation data is available on request.


%

\end{document}